\documentclass[aps, a4paper, amsfonts, amssymb, amsmath, reprint, showkeys, nofootinbib, twoside]{revtex4-1}
\usepackage[english]{babel}
\usepackage[utf8]{inputenc}
\usepackage{textgreek}
\usepackage[colorinlistoftodos, color=green40, prependcaption]{todonotes}
\usepackage{amsthm}
\usepackage{xcolor}
\usepackage{graphicx}
\usepackage{microtype}
\usepackage{cases}
\usepackage[T1]{fontenc}

\usepackage[pdftex, pdftitle={Article}, pdfauthor={Author}]{hyperref} 
\usepackage[normalem]{ulem}
\usepackage{scalerel}
\usepackage{tikz}
\usetikzlibrary{svg.path}

\definecolor{orcidlogocol}{HTML}{A6CE39}
\tikzset{
  orcidlogo/.pic={
    \fill[orcidlogocol] svg{M256,128c0,70.7-57.3,128-128,128C57.3,256,0,198.7,0,128C0,57.3,57.3,0,128,0C198.7,0,256,57.3,256,128z};
    \fill[white] svg{M86.3,186.2H70.9V79.1h15.4v48.4V186.2z}
                 svg{M108.9,79.1h41.6c39.6,0,57,28.3,57,53.6c0,27.5-21.5,53.6-56.8,53.6h-41.8V79.1z M124.3,172.4h24.5c34.9,0,42.9-26.5,42.9-39.7c0-21.5-13.7-39.7-43.7-39.7h-23.7V172.4z}
                 svg{M88.7,56.8c0,5.5-4.5,10.1-10.1,10.1c-5.6,0-10.1-4.6-10.1-10.1c0-5.6,4.5-10.1,10.1-10.1C84.2,46.7,88.7,51.3,88.7,56.8z};
  }
}

\newcommand\orcidicon[1]{\href{https://orcid.org/#1}{\mbox{\scalerel*{
\begin{tikzpicture}[yscale=-1,transform shape]
\pic{orcidlogo};
\end{tikzpicture}
}{|}}}}

\begin{document}
\title{Thermodynamic constraints and future singularities in Unimodular \\
       Gravity driven by phantom and non-phantom fluids}

\author{Norman Cruz$^{1,2}$\orcidicon{0000-0002-0737-3497}}
\email{norman.cruz@usach.cl}

\author{Samuel Lepe$^{3}$\orcidicon{0000-0002-3464-8337}}
\email{samuel.lepe@pucv.cl}

\author{Guillermo Palma$^1$\orcidicon{0000-0001-7326-964X}}
\thanks{corresponding author: guillermo.palma@usach.cl}

\author{Miguel Cruz$^4$\orcidicon{0000-0003-3826-1321}}
\email{miguelcruz02@uv.mx}

\affiliation{$^1$Departamento de F\'{\i}sica, Universidad de Santiago de Chile, Avenida Ecuador 3493, Santiago, Chile,\\
$^2$Center for Inter-disciplinary Research in Astrophysics and Space Exploration (CIRAS), Universidad de Santiago de Chile, Av. Libertador Bernardo O’Higgins 3363, Estaci\'on Central, Chile,\\
$^3$Instituto de F\'\i sica, Pontificia Universidad Cat\'olica de Valpara\'\i so, Casilla 4950, Valpara\'\i so, Chile,\\
$^4$Facultad de F\'{\i}sica, Universidad Veracruzana 91097, Xalapa, Veracruz, M\'exico.}

\date{\today} 

\begin{abstract}
This work investigates future cosmological singularities in a flat FLRW universe filled with a single barotropic fluid, ( $p = (\gamma - 1)\rho$), within the framework of unimodular gravity. In this setting, the non-conservation of the energy–momentum tensor is encoded through an energy diffusion function $Q$. While a constant diffusion term leads to an effective cosmological constant and preserves adiabatic evolution, a time-dependent $Q(t) $ induces non-adiabatic dynamics.
We consider a power-law Ansatz for $Q $ as a function of the redshift and impose the condition of positive entropy production. This requirement leads to non-trivial constraints on the model parameters, with direct implications for the admissible singularity structure. In particular, within the thermodynamically allowed sector, we show that Big Rip singularities are excluded for non-phantom fluids when the cosmological constant is positive.
For phantom fluids, the model reproduces the expected Big Rip behavior, as well as Big Crunch solutions for negative cosmological constant. More importantly, we show that diffusion can induce an effective phantom regime even when the fundamental fluid is non-phantom. In particular, for a negative cosmological constant, we present an explicit realization of a Big Rip singularity in unimodular gravity driven by diffusion, while consistently preserving a non-phantom equation of state and positive entropy production. These results reveal a novel mechanism for the emergence of future singularities, with no direct analogue in standard General Relativity.

\end{abstract}


\maketitle
\section{Introduction}
\label{sec:intro}
Since the formulation of Einstein's general relativity as a geometric theory of gravity, the occurrence of gravitational singularities has remained one of the most profound and challenging problems in theoretical physics, particularly in the area of cosmology, where
the Big Bang singularity is a signal of fundamental incompleteness in our understanding of spacetime geometry and matter behavior at extreme conditions of density and curvature.
Interestingly, since the discovery of the accelerated expansion of the universe and a dark energy (DE) component responsible for this dynamical behavior, the $\Lambda$CDM model has kept its suitability to describe the cosmological data with the simplest form of a dark energy component modeled by the cosmological constant (CC) \cite{Blanchard_2024,efstathiou2025challenges}. However, recent cosmological observations from DESI 2024 indicate a preference for evolving models of DE, which have motivated to constrain different cosmological models of dynamical DE\cite{bousis2024hubble,yang2024quintom, giare2024interacting,colgain2024does,park2024using}. 
 
Among other models of dynamical DE in the framework of general relativity, in recent decades unimodular gravity (UG) naturally leads to variable CC. Moreover, UG has emerged as a particularly compelling framework that is able to address several long-standing conceptual issues in gravitational physics, including the CC problem and the nature of diffeomorphism invariance. Previous investigations in this framework have mainly focused on the cosmological application to the CC problem  \cite{10.1063/1.529283, 10.1063/1.1328077}. However, the implications for spacetime singularities have received considerably less attention, despite their fundamental importance in understanding the full physical consequences of this theory.

Our aim in this paper is to investigate cosmological singularities within the UG framework, in order to assess how their behavior and occurrence may differ from those predicted in standard General Relativity. The possibility of modifications of the energy-momentum conservation laws in this theory opens the landscape of new equations of state of DE, which could drive future singularities, or even how this non-conservation can prevent them.  Of course, the results found are dependent on the energy diffusion function (EDF) chosen to take into account the non-conservation of the cosmic fluids. A physically motivated choice for the EDF is inspired by the \textit{continuous spontaneous localization (CSL)} model~\cite{Pearle:1976ka,Ghirardi:1985mt,Pearle:1988uh,Ghirardi:1989cn}, in which energy is continuously generated via spontaneous quantum collapse events. Previous investigations to address cosmological problems in UG have chosen instead a phenomenologically motivated EDF described by
\begin{equation}
    Q:=\alpha \rho, \label{eq:alfa} 
\end{equation}
where $\rho$ is the energy density of the fluid used in the non conservation equation, and $\alpha$ is a constant. This is  one of the most simple assumptions about this term, which was introduced in \cite{Corral:2020lxt}. It should be noted that diffusion has also been shown to eliminate the emergence of higher-order field equations in cosmological settings \cite{PhysRevD.99.123525}, thereby mitigating potential issues related to the well-posedness of the corresponding Cauchy initial-value problem. In addition, this model follows the behavior of the $\Lambda$CDM model very closely and was used to alleviate the Hubble tension in the framework of interacting fluids \cite{perez2021resolving, LinaresCedeno:2020uxx}. On the other hand, the process of energy diffusion in UG leads to an effective CC that drives the cosmic acceleration, but besides, opened the possibility of an effective phantom behavior for this effective  In the case of $H_0$ tension, a dynamical DE with $\omega < -1$ at low redshifts is favored by cosmological data~\cite{Riess_2018,2019NatAs...3..272R}. In this sense, UG naturally gives rise to novel phantom-like behaviors, thereby broadening the landscape of possible future singularities.

When the thermodynamic behavior of these models is taken into account—where non-adiabaticity naturally arises—previous analyzes have shown that the second law of thermodynamics imposes non-trivial constraints on the admissible EDFs. As demonstrated in \cite{cruz_exploring_2024}, the EDF given by Eq. (\ref{eq:alfa}) exhibits thermodynamic inconsistencies. To ensure compatibility with thermodynamic requirements, in this work we introduce a new EDF, expressed as a function of the redshift, which guaranties a positive growth of cosmic entropy provided certain parameter constraints are satisfied.

In exploring cosmological singularities within the UG framework, we find that the non-conservation of the energy–momentum tensor implies that the evolution of the energy density $\rho$ depends explicitly on the functional form of $Q$. Ensuring the positivity of $\rho$ imposes stringent constraints on the parameters defining $Q$, as well as on the equation of state adopted for the cosmic fluid. In this first exploration, we assume a barotropic EoS for the energy density and pressure of the fluid, i.e., $p=(\gamma-1) \rho$, where $\gamma$ is a constant parameter. Since the study of future singularities typically focuses on the late-time regime—where the cosmic expansion is dominated by dark energy—analyzes often involve a single dark fluid characterized by a specific equation of state, usually of phantom type. In the context of UG, however, we also investigate the possibility that a non-phantom fluid may drive future singularities, as the diffusion function can induce an effective equation of state that mimics phantom behavior, as will be shown later. 

The paper is organized as follows. In Sect. II, we review and summarize the principal features of unimodular gravity. In Sect. III, we explore the thermodynamics requirements that an energy diffusion function $Q$ must satisfy and propose henceforth a suitable Ansatz. In Sect. IV, we find future cosmological singularities as a consequence of the proposed Ansatz, including only those scenarios where the energy density remains positive.  ($8\pi G = c = 1$ units will be used throughout this work).

\section{Background dynamics of Unimodular gravity}
\label{sec:diff}

As already known from some different formulations for UG, see, for instance, Refs. \cite{Einstein:1919gv, BUCHMULLER1988292, PhysRevD.40.1048, 10.1063/1.529283, 10.1063/1.1328077}, the gravitational field is described by a subset of the Einstein field equations.  This subset, comprising the trace-free Einstein equations, permits the reconstruction of the full Einstein equations plus an integration constant. Within the UG description, the dynamical equations for the gravitational field are given by
\begin{equation}
R_{\mu \nu} - \frac{1}{4}g_{\mu \nu}R  = T_{\mu \nu}- \frac{1}{4}g_{\mu \nu}T, 
\label{eq:unieom}
\end{equation}
being $T$ the trace of the energy-momentum tensor of the matter fields, defined as usual. The theory described by the field equations given in (\ref{eq:unieom}) has been dubbed as {\it {unimodular gravity}}. A realization of the field equations given above is provided by introducing the unimodular condition to the Einstein-Hilbert action by means of a Lagrange multiplier $\lambda(x)$; then the Lagrangian density of the gravitational sector takes the form $\mathcal{L} = \sqrt{-g}[R + \lambda(x)(1-\zeta(x)/\sqrt{-g})]$, $\zeta(x)$ being a scalar density. Notice that the variation with respect to the metric of the action constructed with the aforementioned Lagrangian density and matter fields is simply
\begin{equation}
    R_{\mu \nu} - \frac{1}{2}g_{\mu \nu}R + \lambda(x) g_{\mu \nu}  = T_{\mu \nu}, \label{eq:unieom1}
\end{equation}
and the volume element is fixed through the condition $\zeta(x) = \sqrt{-g}$, which is obtained by performing the variation of the action with respect to $\lambda(x)$. From the trace of equation (\ref{eq:unieom1}) one gets
\begin{equation}
    \lambda(x) = \frac{1}{4}(R+T), 
    \label{eq:lagrange}
\end{equation}
which allows us to write (\ref{eq:unieom}) once we replace it back into (\ref{eq:unieom1}).
If we compute the divergence of equation (\ref{eq:unieom1}) with the Lagrange multiplier written in (\ref{eq:lagrange}), due to the Bianchi identity $\nabla^{\mu}G_{\mu \nu}=0$, being $G_{\mu \nu}:= R_{\mu \nu} - (1/2)g_{\mu \nu}R$ the Einstein tensor, we obtain the following
\begin{equation}
    \frac{1}{4}\nabla^{\mu}g_{\mu \nu}(R+T)=\nabla^{\mu}T_{\mu \nu}, \ \ \ \mbox{or} \ \ \ \nabla_{\nu}\lambda(x) = \nabla^{\mu}T_{\mu \nu}.
\end{equation}
The above equation can be solved as
\begin{equation}
    \lambda(x) = \Lambda + Q, \label{eq:diff}
\end{equation}
where $\Lambda$ is an arbitrary integration constant whose value can be fixed by observations, and $Q(x)$ is an arbitrary function that quantifies the violation of the energy-momentum tensor and is defined as $Q(x):=\int_{l} J(x)$, with $J_{\nu} := \nabla^{\mu}T_{\mu \nu}$ being the current for the non-conservation of the energy-momentum tensor. Due to this characteristic, the function $Q$ is generally known as {\it diffusion term}. It is worth mentioning that if the conservation of $T_{\mu \nu}$ is assumed, then $\lambda(x)$ represents a genuine cosmological constant \cite{Ellis_2011, ellis_trace-free_2013}, whereas if $Q$ is not constant, diffusion effects cause $\lambda(x)$ to depart from this constant behavior. 

In this way the field's equation takes the form
\begin{align}\label{eq:fried1}
3H^2 &= \left(\rho + Q \right) + \Lambda ,\\
 \label{eq:accel} 2\dot{H} + 3H^2 &= - \left(p-Q \right) + \Lambda ,
\end{align} 
where a dot denotes the derivative with respect to cosmic time and $H(t)=\dot{a}/a$ is the Hubble function. As $\Lambda$ is an integration constant, one is free to choose the three possible values: $\Lambda<0,\Lambda=0$ and $ \Lambda>0$. Since our goal is to investigate the types of singularities that may arise as functions of the model parameters, we will not restrict ourselves to observationally favored values of the fluid EoS, or of the cosmological constant $\Lambda$. Nonetheless, within the context of the observable universe, the value of $\Lambda$ has been constrained by using cosmological data in the UG framework, as shown in Ref.~\cite{Corral:2020lxt}. On the other hand, a negative cosmological constant cannot be ruled out a priori, as it may be relevant to describe certain phases of cosmic evolution \cite{calderon2021negative, akarsu2020graduated, Visinelli_neg_CC_2019}.

The non-homogeneous continuity equation for the energy densities is given by
\begin{equation}
\dot{\rho } + 3H(\rho  + p )= - \dot{Q },\label{eq:nonconsrho}
\end{equation}
where we have considered, $Q=Q(t)$, in order to preserve the cosmological principle. In the case $Q=0$, as commented above, the $\Lambda$CDM model is recovered if $\Lambda \neq 0$. The equation (\ref{eq:nonconsrho}) for a barotropic fluid can be written as follows
\begin{equation}
    \dot{\rho} + 3H\gamma \rho= - \dot{Q},\label{eq:nonconsrho1}
\end{equation}
which can be cast in the standard form,
\begin{equation}
    \dot{\rho} + 3H\gamma_{\mathrm{eff}}\rho= 0,\label{eq:nonconsrho2}
\end{equation}
where the following definition emerged for the effective parameter state
\begin{equation} \label{eq:omegaefective}
    \gamma_{\mathrm{eff}}:= \gamma + \frac{\dot{Q}}{3H\rho}.
\end{equation}
Consequently, as mentioned above, the inclusion of the diffusion term yields a more generalized fluid description, which is distinguished by an effective variable - as opposed to a constant - parameter of state.

\section{Thermodynamics conditions and Ansatz for the energy diffusion function}
\label{sec:thermo}
In this section, we analyze the conditions under which the diffusion function $Q$ leads to a positive entropy production, ($dS/dt >0$). As shown in Ref.~\cite{cruz_exploring_2024} within the framework of unimodular gravity, a time-dependent diffusion term ($\dot{Q} \neq 0$) induces a non-adiabatic evolution at the background level, whereas a constant $Q$ corresponds to an adiabatic process. In what follows, we consider a universe filled by a single barotropic fluid component in a homogeneous FLRW universe, with no dissipative processes such as viscosity, heat flux, or particle creation, the Gibbs relation \cite{Callen:450289}:

\begin{equation}
    TdS = dU+pdV, \label{eq:gibss}
\end{equation}
 and can be written in terms of the diffusion function as follows \cite{cruz_exploring_2024}
\begin{equation}
    T\frac{dS}{dz} =-V\frac{dQ}{dz}, \label{eq:entropy}
\end{equation}
fully characterizes the thermodynamics of the system. In the above expression, $U$ is the internal energy defined as $U:=\rho V$, $V = V_{0}(1+z)^{-3}$ is the volume, and $T$ the temperature of the cosmic fluid. This result shows that the non-adiabaticity of the system—and therefore the entropy production—is entirely governed by a non-constant diffusion function $Q(t)$. Since no additional fluid components with independent thermodynamic degrees of freedom are introduced, the entropy production is entirely associated with the energy exchange encoded in $Q$, rather than with separate dissipative processes.
 From now on, derivatives w.r.t. time are converted into derivatives w.r.t. the cosmological redshift by means of the standard change of variable, $1+z=a(t)^{-1}$. Notice that in the case of $Q=\mbox{constant}$, the cosmic expansion obeys an adiabatic process, just as the cosmological standard model describes. Due to the change of variable from cosmic time to redshift, the third law of thermodynamics is obeyed as long as
\begin{equation}
    \frac{dQ}{dz} > 0, \ \mbox{implying}, \ \frac{dS}{dz} < 0,\label{eq:good}
\end{equation}
as can be seen from Eq. (\ref{eq:entropy}). It should be noted that the above condition implies that $\frac{dQ}{dt} < 0$, then from Eq.(\ref{eq:nonconsrho1}) we can infer that the thermodynamic second law in UG restricts the EDF to act only as a source of energy on the cosmic fluid and not as a sink. Of course, this result is general,  independent of the particular Ansatz chosen. On the other hand, the sign of $Q(t)$ is not restricted and may act in two ways as a dark energy component, via Eq.(\ref{eq:diff}).

Writing Eq. (\ref{eq:nonconsrho}) in terms of the redshift, we obtain the following expression for the energy density 
\begin{equation}
    \rho(z)=\rho_{0}(1+z)^{3\gamma}\left(1-\frac{1}{\rho_{0}}\int^{z}_{0} \frac{dQ}{dz'}\frac{dz'}{(1+z')^{3\gamma)}} \right), \label{eq:eff1}
\end{equation}
which again demands the negativity of the integral as in the case for the entropy;  the zero subscript denotes evaluation of cosmological quantities at present time which is given by $z=0$, i.e., $\rho_{0} = \rho(z=0)$. 
Let us now briefly outline the general considerations that guided our selection of the Ansatz for the EDF $Q$. In  the case of the barotropic model, $Q=\alpha\rho$, the specific dependence of the EDF on the scale factor $a$ is given implicitly through the dependence on the energy density $\rho (a)$, while for the case of the CSL model for baryons, the energy–momentum violation current leads to $\dot{Q}=-\xi_{\mathrm{CSL}}\rho^{b}$, where $\rho^{b}$ represents the energy density of the baryonic fluid, and $\xi_{\mathrm{CSL}}$ is an experimentally constrained parameter that determines the sign of $Q$ \cite{perez2019dark}.

Alternatively, our choice is motivated by two key considerations: i) the thermodynamic condition given by Eq. (15) requires that suited EDF must be a monotonically increasing functions of the redshift; ii) a natural generalization Ansatz for $Q(a)$ should encompass earlier functional dependence models. A particularly prominent class of models is obtained by assuming a power-law dependence on the redshift, whose scaling differs from that of the energy density. 

Beyond the arguments presented above, the framework considered here is essentially phenomenological, with the second law of thermodynamics providing the primary physical constraint on the diffusion function. Hence, we adopt the following Ansatz in terms of the redshift:

\begin{equation}  \label{Ansatz}
Q\left( z\right) = Q_{0} \left( 1+z\right) ^{\beta},
\end{equation}
where $Q_{0}$ has the same dimensions as $\rho_{0}$.  From this equation, we obtain the following.
\begin{equation}  \label{dQdz}
\frac{dQ (z)}{dz}=\beta Q_{0}\left(
1+z\right) ^{\beta -1}.
\end{equation} 

Condition $dQ/dz>0$ gives two possible scenarios: i) $ Q_{0}>0, \beta >0$ or ii) $ Q_{0}<0, \beta <0$, i.e.,
\begin{subnumcases}
  {
    Q(z) =
  }
  Q_{0}(1+z)^{\beta} \Rightarrow Q_{0}>0,\ \beta > 0,
  \label{eq:term1} \\
  -\frac{\left|Q_{0}\right|}{(1+z)^{\left|\beta \right|}} \Rightarrow \ Q_{0}<0,\ \beta < 0.  \label{eq:term2}
\end{subnumcases}

Notice that the contribution from $Q$ to the non-conservation of energy-momentum may, in principle, increase or decrease during cosmic evolution. 

Inserting the expression for the diffusion term given by Eq.(\ref{Ansatz}) in Eq. (\ref{eq:eff1}) allows us to obtain the energy density in terms of the redshift $z$
\begin{eqnarray}
    &\rho(z)& = \rho_{0}(1+z)^{3\gamma}\times \nonumber \\ 
   & \times & \left(1+\frac{\beta Q_{0}}{\rho_{0}(\beta - 3\gamma)} \left[1-(1+z)^{\beta-3\gamma}\right]\right). \label{eq:energy1}
\end{eqnarray}

In order to discuss the effective EoS of the cosmic fluid for the chosen $Q(z)$, we rewrite Eq.(\ref{eq:nonconsrho1}) in terms of the redshift by using the identity $dz/dt=-(1+z)H$, which leads to  
\begin{equation}
    \frac{d\rho}{dz} -\frac{3}{1+z} \biggl[ \left( \gamma - \frac{1+z}{3\rho}\frac{dQ}{dz}\right )\rho \biggr] =0,\label{nonconsrhodez}
\end{equation}
which implies an effective EoS given by
\begin{equation}
    \gamma_{eff} = \gamma - \frac{1+z}{3\rho}\frac{dQ}{dz}\label{gammaeffdez}.
\end{equation}
Observe that the thermodynamic requirement $\frac{dQ}{dz}>0$ in Eq. (\ref{gammaeffdez}) indicates, within the UG framework, that the fluid assumed to fill the universe (a single component in our setup) must exhibit an effective EoS with pressure lower than its intrinsic one. Consequently, if the dark matter component is pressureless, it effectively behaves as a quintessence-type fluid. 
 
\section{Cosmological singularities}
\label{sec:sing}
In this section, we will study the existence of different types of future curvature singularities in the framework of UG gravity, using the diffusion function $Q$ . In what follows, we shall adopt the classifications given in \cite{tiposdeBigRip,clasificacionBigRipdetallada}:

\begin{itemize}
\item \textbf{Type 0A}  (``Big Bang''): for $t\rightarrow 0$ , $a \rightarrow 0$, $\rho \rightarrow \infty$ and $|p|\rightarrow \infty$. In terms of the Hubble parameter: $H \rightarrow \infty$ and also its cosmic time derivatives.
 \item \textbf{Type 0B} (``Big Crunch''): for $t\rightarrow t_{s}$ , $a \rightarrow 0$, $\rho \rightarrow \infty$ and $|p|\rightarrow \infty$. In terms of the Hubble parameter: $H \rightarrow \infty$ and also its cosmic time derivatives. In the framework of GR this singularity is usually
 present in a closed universe and filled with matter
 satisfying the strong energy conditions, or a negative cosmological constant.
 \item\textbf{Type I}  (``Big Rip''): for $t\rightarrow t_{s} < \infty$, $a \rightarrow \infty$, $\rho \rightarrow \infty$, and $|p|\rightarrow \infty$. At time $t = t_{s}$, the Hubble parameter and its cosmic time derivative also diverge \cite{caldwell2003phantom, gonzalez2004k}. This singularity occurs when the universe is filled with a phantom dark energy fluid; nevertheless, in the framework of Eckart's theory a dissipative fluid with a barotropic EoS could behave like an effective phantom EoS, driving also this singularity \cite{cruz2022singularities}. 
\item\textbf{Type $\mathbf{I_{l}}$}  (``Little-Rip''): for $t\rightarrow \infty$, $a \rightarrow \infty$, $\rho \rightarrow \infty$ and $|p|\rightarrow \infty$. At $t\rightarrow \infty$ the Hubble rate and its cosmic time derivative also diverge.\cite{frampton2011little, FRAMPTON2012204,brevik2011viscous, bouhmadi2013tradeoff, albarran2016classical}. 
\item \textbf{Type II}  (``Sudden''): for $t\rightarrow t_{s}$, $a \rightarrow a_{s}$, $\rho \rightarrow \rho_{s}$ and $|p|\rightarrow \infty$. 
\item \textbf{Type III} (``Big freeze'') : for $t\rightarrow t_{s}$, $a \rightarrow a_{s}$, $\rho \rightarrow \infty$, and $|p|\rightarrow \infty$. 
\item \textbf{Type IV} (``Generalized Sudden''): for $t\rightarrow t_{s}$, $a \rightarrow a_{s}$, $\rho \rightarrow 0$ and $|p|\rightarrow 0$, and higher order derivatives of $H$ diverge.
\end{itemize}
The Ricci scalar curvature of a cosmological flat FLRW model is given by the following expression 
\begin{equation}\label{ricci}
    R=6\left({\frac {{\ddot {a}}(t)}{a(t)}}+{\frac {{\dot {a}}^{2}(t)}{a^{2}(t)}}\right)= 6\left(\dot{H}+2H^{2}\right), 
\end{equation}
which means that divergences in the parameters $H$ or $\dot{H}$, associated to a particular cosmological solution, also implies divergence in this term.

\subsection{Characterization of the physical parameter region of the model}
\label{sec:sing2}

The aim of this subsection is to characterize the region of the model-parameter space for which the physics is well-settled, that is, where the non-negativity of the dark matter density $\rho$, the existence of a well-defined Hubble parameter ($H^2(a) \geq 0$), and the positivity of entropy production are guaranteed. 

Notice that the energy density of the fluid given by Eqs.(\ref{eq:energy1}) can be rewritten as a sum of two terms
\begin{equation}
     \rho(z) = \rho_{0}(1+z)^{3\gamma}\left(1+\frac{Q_{0}\beta}{\rho_{0}(\beta - 3\gamma)} \right)-\frac{Q_{0} \beta}{\beta-3\gamma} (1+z)^{\beta} \label{eq:energy3}, 
\end{equation}
which means that the EDF introduces a significant modification on the evolution of the main fluid that filled the universe in terms of the redshift, and clarify the form of the parameter $B$ defined below. Demanding that $\rho >0$ in Eq. (\ref{eq:energy3}) with $\gamma>0$ the fluid will satisfy the weak energy condition, i.e., $\rho\geq 0$ and $\rho+p\geq 0$, since we adopted a barotropic EoS, which results in $\left|p\right|\leq \rho$. It will not be the case if $\gamma<0$ (phantom fluid), We consider the expression given by Eq. (\ref{eq:energy3}) as true expression of the fluid under the EDF $Q(t)$ since is the solution to the non conservation equation.  In this sense, in Eq. (\ref{eq:fried1}) the term $\Lambda +Q(t)=\lambda (t)$ represents an effective CC and the requirement of positivity is out of place.  

In what follows, we will work with the scale factor $a(t)$ instead of the redshift in order to evaluate the behavior of this parameter in terms of cosmic time and its possible future singularities. We first start by performing a direct integration of Eq.~(\ref{eq:nonconsrho}) using the variation of the homogeneous solution method, which yields the following expression for the matter energy density:
\begin{equation}
\rho \left(a\right) = A ~ a ^{-3\gamma} + B~ a ^{-\beta},
\label{rho_m_a}
\end{equation}
with $A$ and $B$ defined, respectively, by 
\begin{equation}
A = \rho_{0} ~\left(1 - \frac{\beta Q_{0} / \rho_{0}}{3\gamma -\beta} \right) \quad \text{and} \quad B = \frac{\beta ~ Q_{0}}{3\gamma -\beta} .
\label{def_A_B}
\end{equation}

The second relevant constraint comes from Eq.~(\ref{eq:fried1}), and can be expressed as

\begin{equation}
H^2(a)= \left[\frac{A}{3} ~ a^{-3\gamma} + \frac{\Lambda}{3}+ \tilde{B}~ a^{-\beta} \right] \geq 0,
\label{Hubble_2_a}
\end{equation}
with $\tilde{B} = B~ \gamma /\beta$. 

Finally, the last constraint on the model parameters is provided by the positivity of entropy production discussed below Eq.~(\ref{dQdz}), which can be recast as:
\begin{equation}
\beta ~ Q_{0} > 0.
\label{thermo_law}
\end{equation}
To ensure that the matter energy density remains positive, it is necessary to impose certain constraints on the model parameters. We therefore classify the cases according to whether $\gamma$ takes positive or negative values:

\subsubsection{Non -phantom fluid (\texorpdfstring{$\gamma > 0$}{gamma > 0}).}
\label{sec:phantom0}

 We distinguish two cases  
    \begin{itemize}
            \item $\beta >0$. This implies that $Q_{0} >0$ due to the positiveness of the entropy production. There are two further    possibilities:
                 \begin{itemize}
                       \item  $3\gamma > \beta $, which constrains the first coefficient of $\rho$ to be non-negative, since the early-time regime is dominated by the first contribution, that is, $\rho_{0} \geq \beta~ Q_{0} / (3\gamma - \beta)$.  
                       \item $3\gamma < \beta $, which leads to a negative coefficient $B$. Nevertheless, since the early-time period is dominated precisely by $B~a^{-\beta}$ (the second contribution in Eq.~(\ref{rho_m_a})), this parameter region must be excluded.
                 \end{itemize}
Now, we include in our analysis the existence of a well-defined $H(a)$. We just need to consider only the case $3\gamma > \beta > 0$, together with the validity of inequality $\rho_{0} \geq \beta~ Q_{0} / (3\gamma - \beta)$. Since both coefficients $A$ and $\tilde{B}$ are positive in the entire region, for $\Lambda > 0$ the existence of a real Hubble parameter is guaranteed, while for $\Lambda < 0$, its existence requires a lower bound for the scale parameter.       
          \item $\beta <0$. This implies that $Q_{0} < 0$ due to the positiveness of the entropy production. The late-times regime is dominated by the second and third contributions appearing in $\rho$ and $H(a)$ (see Eqs.~(\ref{rho_m_a})-(\ref{Hubble_2_a})). Since $B > 0$, while $\tilde{B}= B \gamma/\beta < 0$, it is not possible to guaranty the existence of the Hubble parameter beyond an upper bound for the scale parameter.  We will address this subtle point later on.             
    \end{itemize}
      
\subsubsection{Phantom fluid (\texorpdfstring{$\gamma < 0$}{gamma < 0}).}
\label{sec:phantom}

As in the former case, we distinguish two cases:  
\begin{itemize}
            \item $\beta < 0$. This implies that $Q_{0} < 0$ due to the positiveness of the entropy production. There are two further    possibilities:
                 \begin{itemize}
                       \item  $3\gamma > \beta $, which constrains the first coefficient of $\rho$ to be non-negative, since the early-time regime is dominated by the first contribution, that is, $\rho_{0} \geq \beta~ Q_{0} / (3\gamma - \beta)$.  
 
                       \item $3\gamma < \beta $, which leads to a negative coefficient $B$. Nevertheless, since the late-time region is dominated precisely by $B~a^{-\beta}$ (the second contribution in Eq.~(\ref{rho_m_a})), this parameter region must be excluded, or an upper bound on $a$ must exist.  \\
                 \end{itemize}

                 Now, the existence condition on the Hubble parameter $H(a)$ is taken into account. We just need to consider only the case $0 > 3\gamma > \beta $, together with the validity of inequality $\rho_{0} \geq \beta~ Q_{0} / (3\gamma - \beta)$. Since both coefficients $A$ and $\tilde{B}$ are positive quantities within this region, for $\Lambda > 0$ the existence of a real Hubble parameter is guaranteed, while for $\Lambda < 0$, its existence requires a lower bound for the scale parameter.\\
                 
            \item $\beta > 0$. This implies that $Q_{0} > 0$ due to the positiveness of the entropy production. The coefficient $A$ must be positive, since the first contribution to $\rho$ is $Aa^{\beta}$, and it dominates the large-time behavior. This is achieved by demanding $\rho_{0} \geq \beta~ Q_{0} / (3\gamma - \beta)$. At early times, the energy density $\rho$ is dominated by the second term (see, for instance, Eq.~(\ref{rho_m_a})), however, the corresponding coefficient $B$ is negative. Consequently, a lower bound on the scale factor $a$ must exist in order to ensure the non-negativity of $\rho$. This issue will be examined in detail in the next subsection.
                 
\end{itemize}

A comment on the particular case \underline{$3\gamma = \beta $} is in order. Starting from the continuity equation for the matter energy density, (see Eq.~(\ref{eq:nonconsrho1}), one obtains the explicit analytical expression

\begin{equation}
\rho(a) = \biggl[ \rho_{0} + 3\gamma Q_{0} ~ln (a) \biggr] ~a^{-3 \gamma}.
\label{rho_a_3gamma_beta}
\end{equation}

This expression becomes negative at early times, i.e., for $a < exp {(-\rho_{0}~/ 3~\beta ~ Q_{0})}$, which is inconsistent with the well-established behavior of the matter energy density in the early universe. Consequently, the case $3\gamma = \beta $ must be excluded, as it makes the above expression ill-defined and leads to an unphysical negative energy density $\rho (a)$ at early-times. 

\subsection{General Analysis of Singularities}
\label{sec:ge_singular}

An important consequence of the above constraints is that future cosmological singularities with positive $\Lambda$, such as Big Rip divergences at finite time-value, cannot be induced solely by the energy diffusion function in a universe devoid of phantom fluid components. 

To analyze this possibility, we use the constraint of Eq.~(\ref{eq:fried1}) 

\begin{equation}
t-t_{*} = \int^{a}_{a_*} \frac{ da / a}{ H(a)} 
\label{scale_parameter}
\end{equation}
where the Hubble parameter $H(a)$ is defined by
\begin{equation}
H(a)= \left[\frac{A}{3} ~ a^{-3\gamma} + \frac{\Lambda}{3} + B \frac{\gamma}{\beta}~ a^{-\beta} \right]^{1/2},
\label{Hubble_a}
\end{equation}
with $A$, $B$ given by Eq.~(\ref{def_A_B}), and for $\gamma > 0$.

\subsubsection{Non phantom fluid, \texorpdfstring{$\gamma > 0$}{gamma > 0}, and singularities}
\label{sec:phantom1}
\textbf{Case $\Lambda>0$}. From condition $\gamma > 0$, it follows that the parameter region must satisfy $3\gamma > \beta$. It means that the leading contributions in the large $a$ regime for $\beta > 0$ correspond to $\Lambda/3 + \tilde{B} a^{-\beta}$, and therefore, the asymptotic behavior of the above integral, i.e. for large scale-parameter $a$ values, is given by the expression

\begin{eqnarray}
t-t_{*} &=& \frac{2}{\beta} \sqrt{\frac{3}{\Lambda}} \tanh^{-1} \left[\sqrt{1 + \frac{3 \gamma}{\beta\Lambda} ~a^{-\beta}}~ \right] \nonumber - \\
        & & \frac{2}{\beta} \sqrt{\frac{3}{\Lambda}} \tanh^{-1} \left[\sqrt{1 + \frac{3 \gamma}{\beta\Lambda} ~a_{*}^{-\beta}}~ \right]
\label{no_Big_Rip}
\end{eqnarray}
where $a_{*} = a(t_{*})$, and $t_{*}$ is some particular cosmological value within the validity of the asymptotic expansion, i.e., $a_{*} \gg 1 $. 

From the above expression, we conclude that the limit $a \rightarrow \infty$, necessarily implies $\Delta t \rightarrow \infty$,  since  $\tanh^{-1}(x)$ diverges at $x=1$. In other words, no Big Rip singularity exists in the present parameter region, when the  matter component does not behave as a phantom fluid.

Nevertheless, according to the argument above, the scale factor exhibits an asymptotic divergence as $t \rightarrow \infty$. This late-time behavior corresponds to the de Sitter solution, characterized by $ H(a) \rightarrow  \sqrt{\Lambda / 3}$, together with $\rho \rightarrow 0$ and $\dot{H} \rightarrow 0$. \\
We can conclude that for positive $\Lambda$, and when the matter component behaves as a conventional fluid with $\gamma \approx 1$, the energy diffusion function $Q$ alone is insufficient to induce a Big Rip singularity and the usual de Sitter late time behavior is obtained.

\textbf{Case $\Lambda<0$}. \textbf{Big crunch singularity}. We investigate the consequences of a negative cosmological constant, concentrating on the circumstances that could cause the universe to recollapse and end in a Big Crunch. To illustrate this scenario, we consider the case $\gamma > 0$ (non-phantom fluid), $\beta > 0$, together with $3\gamma > \beta$. The Hubble parameter in this parameter region is

\begin{equation}
H(a)= \left[\frac{A}{3} ~ a^{-3\gamma} - \frac{|\Lambda|}{3}+ \tilde{B}~ a^{-\beta} \right]^{1/2},
\label{Hubble_Lambda_neg}
\end{equation}
must be non-negative, leading to an upper bound on the scale parameter $a_{max}$. This value can be numerically determined for some given physically allowed $\gamma$ and $\beta$, or can be perturbatively found as follows. For a sufficiently small value of the upper bound $a_{max}$, the leading contribution to $H(a)$ is the first. More precisely, for 

\begin{equation}
\frac{A}{3 \tilde{B}}  \gg ~a_{max}^{3\gamma - \beta}, 
\label{succ_aprox}
\end{equation}
the upper bound, up to first order, is given by 

\begin{equation}
a_{ub}^{I} \approx \left( \frac{ A}{\left| \Lambda \right|} \right) ^{ 1 / 3\gamma }.    
\end{equation}

For smaller $a$-values, Eq.~(\ref{succ_aprox}) remains valid, and hence we can approximately compute the integral of Eq.~(\ref{scale_parameter}) obtaining up to leading order 
\begin{equation}
\Delta t \approx \sqrt{3} \int^{a} \frac{ da / a}{\left( A ~ a^{-3\gamma} - \left| \Lambda \right| \right)^{1/2}}  
\label{asymp_big_crunch}
\end{equation}
Using a change of variables $a^{3\gamma} = ( A / \left| \Lambda \right|) \sin^2 ( \theta/2)$, the result can be cast as

\begin{equation}
\Delta t  \approx \frac{ 1}{ \sqrt{3} \gamma \left| \Lambda \right|^{1/2}} ~ \theta,  
\label{leading_order_Hubble}
\end{equation}
which leads to the solution;

\begin{equation}
a^{I}(t) \approx \left[ \frac{ A}{ \left| \Lambda \right|}~ \sin^2 \left( \frac{\sqrt{3\left| \Lambda \right|}~ \gamma ~t}{2} \right) \right]^{1 / 3 \gamma}.
\label{bouncing_solution}
\end{equation}
with an associated Big-Crunch time 

\begin{equation}
t_{BC}^{I}  \approx \frac {2\pi}{\sqrt{3\left| \Lambda \right|} ~\gamma} .  
\label{Big_Crunch_time}
\end{equation}
This singularity at $a=0$ indeed corresponds to a Big Crunch, since the Hubble parameter, its time derivative, and the matter energy density diverge at $t= T_{BC}$, as can be straightforwardly seen from their corresponding definitions. 

To illustrate how the perturbative framework developed above successfully describes a Big Crunch singularity in the case $\Lambda < 0$, we focus on the specific choice of model parameters $3\gamma = 2$ and $\beta = 1$. To guaranty the existence of the Hubble parameter $H(a)$, the scale parameter must be located in the interval $ [0, ~a_{max}]$, where the upper value is

\begin{equation}
a_{max} = \frac{3\tilde{B}}{2~|\Lambda|} +  \frac{3\sqrt{D}}{2 |\Lambda|}.    
\end{equation}
being $D$ the discriminant $D = \tilde{B}^2 + 4 A |\Lambda| /9$. 

Within this integration region, the integral of Eq.~(\ref{scale_parameter}) can be solved analytically, yielding

\begin{eqnarray}
t - t_{*}  &=& \sqrt{\frac{3}{|\Lambda|}} \sin^{-1} \left[ \frac{2 |\Lambda|}{3\sqrt{D}} \left(a - \frac{3\tilde{B}}{2~|\Lambda|}                  ~\right)\right] - \\ \nonumber  
           & & \sqrt{\frac{3}{|\Lambda|}} \sin^{-1} \left[ \frac{2 |\Lambda|}{3\sqrt{D}} \left(a_{*} - \frac{3\tilde{B}}{2~|\Lambda|}                  ~\right)\right],
\label{exact_a_Lambda_neg}
\end{eqnarray}
which is a well-defined functional relation for $t(a)$ for $a$ lying in its definition domain. Choosing $t_{*}$ such that $a_{*} = a(t_{*}) = 0$, i.e. $ t_{*} = - \sqrt{3 / |\Lambda|} \sin^{-1} (\tilde{B}/ \sqrt{D})$, which exists since $\tilde{B}/ \sqrt{D} < 1$, the above expression can be inverted to obtain the scale parameter $a$ as a function of the cosmological time

\begin{equation}
a(t) = \frac{3\tilde{B}}{2~|\Lambda|} +  \frac{3\sqrt{D}}{2 |\Lambda|}~\sin \left( \sqrt{\frac{|\Lambda|}{3}} ~t \right). 
\end{equation}
Since $a(t)$ is a monotonically increasing function, its maximum lies when the argument is $\pi/2$, that is, starting from $t_{*}$ the total time to reach the first maximum is

\begin{equation}
t_{BC} = \sqrt{\frac{3}{|\Lambda|}} \frac{\pi}{2} + \sqrt{\frac{3}{|\Lambda|}} \sin^{-1} (\tilde{B}/ \sqrt{D}).
\label{BR_time_Lamda_neg_3Gamma_Beta}
\end{equation}
Now, if we consider the same supposition as the general perturbative computation of $t_{BC}$, namely, for $A~|\Lambda| \ll 9 \tilde{B} /4$, the above result goes into 

\begin{equation}
    t_{BC} = \sqrt{\frac{3}{|\Lambda|}} ~\pi,
    \label{BC_time_3Gamma_Beta_Lambda_neg}
\end{equation}
since, up to first order, $\sin^{-1}(\tilde{B}/ \sqrt{D}) \approx \sin^{-1}(1) = \pi / 2$. This expression agrees exactly with Eq.~(\ref{Big_Crunch_time}), corresponding to the general case for $\Lambda < 0$, when the values $3\gamma = 2$ and $\beta= 1$ are inserted.

Higher-order corrections for the Big-Crunch expressions can be straightforwardly computed, assuming that the condition of Eq.~(\ref{succ_aprox}) holds. Instead of doing so, one can alternatively study the integral of Eq.~(\ref{scale_parameter}) by searching for Big Crunch scenarios in the complementary parameter region, defined by

\begin{equation}
\frac{A}{B} \ll a_{tp}^{(3\gamma - \beta)}. 
\label{succ_aprox_II}
\end{equation}

In this case, there is again an approximated turning point defined by
\begin{equation}
a_{tp}^{II} \approx \left( \frac{3 B}{\left| \Lambda \right|} \right) ^{ 1 / \beta }.    
\end{equation}
A similar analytical computation along the lines delighted above leads to the leading-order expression for the bouncing solution

\begin{equation}
a^{II}(t) \approx \left[ \frac{3 B}{ \left| \Lambda \right| }~ \sin^2 \left( \frac{\sqrt{\left| \Lambda \right|/3}~ \beta ~t}{2} \right) \right]^{1 / \beta},
\label{bouncing_solution2}
\end{equation}
with an associated Big-Crunch time 

\begin{equation}
t_{BC}^{II}  \approx \frac {2\pi}{\sqrt{\left| \Lambda \right|/3} ~\beta} .  
\label{Big_Crunch_time_II}
\end{equation} 
The key point we wish to emphasize is that, regardless of the specific parameter region considered in the analysis of the integral in Eq.~(\ref{scale_parameter}), the model admits Big Crunch solutions whenever the cosmological constant takes negative values. 

\textbf{Big rip singularities.} We will show here that future Big Rip singularities may arise, even if the matter component does not behave as a phantom fluid ($\gamma \geq 0$). In order to illustrate the existence of these kinds of scenarios, let us consider the particular case in which $\rho_{0} = \beta~Q_{0} / (3\gamma - \beta)$, with $\beta < 0$, and hence $Q_{0}< 0$ (to ensure the second law of thermodynamics). The integral of Eq.~(\ref{scale_parameter}) goes for $\Lambda < 0$ into

\begin{equation}
t-t_{min} = \int^{a}_{a_{min}} \frac{ da / a}{ \left( -\left| \Lambda \right|/3 + B~ a^{|\beta|} ~\right)^{1/2}} 
\label{Big_Rip_scale_parameter}
\end{equation}\\
where $t_{min}$ is the cosmological time defined by the minimum value of the scale parameter $a_{min} = a(t_{min})$, defined by the condition $H(a_{min}) = 0$, i.e. $a_{min} = (~ |\Lambda| / 3B ~)^{1/|\beta|}$. After some analytic manipulations, it is easy to show that the above integral can be computed analytically, obtaining the following: 

\begin{equation}
t-t_{min} = \frac{2 \sqrt{3}}{\left| \beta \right| \sqrt{\left| \Lambda \right|}} \tan^{-1}\left( \sqrt{-1 + \frac{3 B}{\left| \Lambda \right|}~ a^{\left| \beta \right|}} ~\right), 
\label{Big_Rip_implicit_solution}
\end{equation}
since $\tan^{-1}(0) = 0$. The above expression can be solved for the scale parameter $a$, giving rise to the solution

\begin{equation}
a (t) = a_{min} ~\left[ ~ \sec^2 \left( \sqrt{\frac{\left| \Lambda \right|}{3}} ~\frac{|\beta|}{2}~ (t-t_{min}) \right) \right]^{1 / \left| \beta \right|},  
\label{Big_Rip_solution}
\end{equation}
which becomes singular at particular cosmological time $t = t_{BR}$, given explicitly by 

\begin{equation}
t_{BR}-t_{min} = \frac{\pi}{\left|\beta\right|} \sqrt{\frac{3}{|\Lambda|}} .
\label{BR_time}
\end{equation}
Moreover, since $H(a(t))$ diverges at $t_{BR}$, together with the energy density $\rho$ (see, for instance, Eq.~(\ref{rho_m_a})), we conclude that $t_{BR}$ corresponds to a Type I singularity, as defined in Section~\ref{sec:sing}, that is, a Big Rip singularity.

\subsubsection{Phantom fluids, \texorpdfstring{$\gamma < 0$}{gamma < 0}, and singularities}
\label{sec:phantom2}
\textbf{Case $\Lambda>0$}. In this case, Big Rip scenarios may occur in the region $\gamma < 0$, as can be demonstrated as follows. We begin by introducing the scales $a_{phQ}$ and $a_{ph \Lambda}$ by
\begin{equation}
a_{phQ} = \left( \frac{A}{B} \right)^{\frac{1}{(\left|\beta\right| - 3 \left|\gamma\right|)}}  ~\text{and} \quad a_{ph \Lambda} = \left( \frac{\Lambda}{3A} \right)^{ 1/ 3 \left|\gamma\right|}.
\end{equation}
As discussed above, for $\gamma < 0$, the requirements of a non-negative energy density and a well-defined Hubble parameter, together with the positivity of the entropy production, constrain the model parameters to satisfy the inequality $\beta < 3\gamma$.  Now, for a scale parameter $a$ larger than both scales, ($a \gg \mbox{Max}\{a_{phQ},  a_{ph \Lambda}\}$, follows that the integral of Eq.~(\ref{scale_parameter}) has the asymptotic expansion

\begin{equation}
t-t_{0} \approx \int^{a}_{a_{0}} \frac{ da~ a^{-(1 + \left|\beta\right|/2)}} { \sqrt{B} } \left( 1 -\frac{A}{2B} a^{-\left|\beta - 3 \gamma\right|} + ... \right)
\label{neg_g_and_beta_positive_Lambda_scale_parameter}
\end{equation}
Up to first order in this expansion, one obtains the following
\begin{equation}
t - t_{0}\approx  \frac{2}{ \left|\beta\right| \sqrt{B}} ~  \left( 1 - a^{-\left|\beta\right|/ 2}\right) . 
\label{BR_first_order}
\end{equation}
The above expression enables an explicit determination of the Big Rip time, defined as the time measured from $t_{0}$, at which the scale factor $a$ diverges:
\begin{equation}
t_{BR}^{(1)} - t_{0}\approx  \frac{2}{ \left|\beta\right| \sqrt{B}}.
\label{BR_first_order_2}
\end{equation}
Within this general scenario, it is straightforward to go beyond first order and compute higher contributions to the Big Rip time. We therefore conclude that in the entire region $\beta < 3~ \gamma < 0$, Big Rip singularities arise, with the scale factor $a$ diverging at a finite time $t_{BR}$, which may be obtained approximately via asymptotic expansions for large $a$ using Eq.~(\ref{BR_first_order}), or computed numerically from Eq. (\ref{scale_parameter}), for a particular model defined by given values of $\gamma$, $\Lambda$, and $\beta$.

\textbf{Case $\beta = -2$, $3\gamma = -1$}. In this case, Eq.(\ref{scale_parameter}) becomes
\begin{equation} 
t - t_{0} = \int_{a_0}^{a} \frac{da/a}{\sqrt{Aa + \Lambda / 3 + B a^{2}}},
\label{part_b_g_Lambda}
\end{equation}
where a specific sign for $\Lambda$ is not assumed here. The argument of the square root in the denominator, which corresponds to the square of the Hubble parameter, can be written using the factorization property as 
\begin{equation} 
H^{2}(a) = B (a - a_{+})(a-a_{-})~ \text{with} \quad a_\pm = \frac{-A}{2B} \pm \sqrt{D},
\end{equation}
with the discriminant of the quadratic polynomial given by $D = (A/2B)^2 -\Lambda /3B$. Since $H^2(a)$ in a non-negative function, the discriminant must satisfy condition $D \leq 0$, in the complete physical range of the scale parameter, i.e. $a \geq 0$. For $\Lambda > 0$, we must distinguish three scenarios.

i) Case $A^2 < 4 B \Lambda /3 $~: A straightforward calculation yields the primitive integral in Eq.~(\ref{part_b_g_Lambda}),which—after some algebraic manipulations—leads to the following explicit analytical expression for the scale factor 

\begin{equation}
a(t) = \frac{2 \Lambda/3\sqrt{\Delta}}{\sinh \left(\sinh^{-1}\left[\frac{1+3A/2\Lambda}{3\sqrt{\Delta}/2\Lambda} - (\tau-\tau_{0})\right] \right) - \frac{A}{\sqrt{\Delta}}}
\label{a_for_part_b_g_Lambda}
\end{equation}  
where $\tau = \sqrt{\Lambda /3} ~t$, and $\Delta$ is defined by $\Delta = 3B / \Lambda - (3A/2\Lambda)^2 $, which is a positive quantity and, therefore, $\sqrt{\Delta}$ is a real number. 

Now, since the function $\sinh(x)$ is a surjective function, there always exists a value for $t-t_{0}$ such that the denominator of the above expression vanishes. This value corresponds exactly to the Big Rip singularity, as we will show below. Analytically, this value is given explicitly by the expression
\begin{equation}
t_{BR}-t_{0} = \sqrt{\frac{\Lambda}{3}} \left[ \sinh^{-1} \left( \frac{2\Lambda}{3\sqrt{\Delta}} + \frac{A}{\sqrt{\Delta}} \right) - \sinh^{-1} \left(\frac{A}{\sqrt{\Delta}} \right) \right].
\label{t_BR_for_part_b_g_Lambda}
\end{equation}
The Hubble parameter $H$ is explicitly given as a function of the scale parameter as $H(a) = \sqrt{Aa + \Lambda / 3 + B a^{2}}$, which allows us to conclude that $H (t)$, and its derivative $\dot{H}(t) = Ba^2 + A a /2$, diverge as the singular value $t_{BR}$ is approached.

Furthermore, using Eq.~(\ref{rho_m_a}) for the matter energy density as a function of $a$, together with the equation of state $p = (\gamma -1) \rho $, we find that both quantities diverge at the singular point $t_{BR}$. This behavior corresponds, according to the standard classification of cosmological singularities, to a Type I (Big Rip) singularity.

ii) Case $A^2 = 4 B \Lambda /3 $~: In this case, the Hubble parameter does not vanish at the origin $a=0$. Hence, no Big Bang singularity exists, and the integral of Eq.~(\ref{part_b_g_Lambda}) leads to the expression: 

\begin{equation}
a(t) = \frac{\exp{\left(\sqrt{\Lambda/3}(t-t_{0})\right)}} {1 + 3A/2\Lambda \left[ 1 - \exp{\left(\sqrt{\Lambda/3}(t-t_{0})\right)}\right]}.
\label{a_for_part_b_g_Lambda_D_eq_0}
\end{equation} 
The above equation shows a future singularity when $t = t_{*} < \infty$, where

\begin{equation}
t_{*} = t_{0} + \sqrt{\frac{3}{\Lambda}} ~\ln{\left(1 + \frac{2\Lambda}{3A}\right)}.
\label{a_sing_D_eq_0}
\end{equation}
As in the previous case, the Hubble parameter diverges at $t=t_{*}$, and consequently $H(t)$ also becomes unbounded, \[ \lim_{t \rightarrow t_{*}} H(t) = \infty \]Similarly, $\dot{H}(t)$, the matter energy density $\rho$ and the pressure $p$ diverge at $t = t_{*}$, leading to the conclusion that this time corresponds to a Big Rip singularity.

iii) Case $A^2 > 4 B \Lambda /3 $~: For a positive discriminant, the integral of Eq.~(\ref{part_b_g_Lambda}) goes into the implicit expression

\begin{equation}
G(a(t)) \exp{\left(\sqrt{\frac{\Lambda}{3}}(t-t_{0})\right)} = G (a(t_{0}),
\label{a_for_part_b_g_Lambda_D_greater_0}
\end{equation} 
where 
\begin{equation}
G(a(t)) = \frac{2}{a} + \frac{3A}{\Lambda} + 2 \sqrt{\frac{3B}{\Lambda} + \frac{3A}{\Lambda} \frac{1}{a} + \frac{1}{a^2}}
\label{G_of_a_for_part_b_g_Lambda_D_greater_0}
\end{equation} 
is a monotonically decreasing function of $a$, i.e. fulfilling $G(a(t)) < G(1)$ for $a({t_{0}})= 1$ and $t > t_{0}$. Hence, Eq.~(\ref{a_for_part_b_g_Lambda_D_greater_0}) can always be solved for $t_{*} - t_{0}$ for a given scale parameter $a_{*} > 1$, because of the surjective property of the exponential function in the physical domain $[0,\infty)$. 

In particular, for $a \rightarrow \infty$, $ G(a \rightarrow \infty)\rightarrow  3~A / \Lambda + 2 \sqrt{3~B /\Lambda}$, which is a positive quantity. As a consequence, there exists a singularity at the finite time $t_{*}$ given by

\begin{equation}
t_{*} = t_{0} + \sqrt{\frac{3}{\Lambda}} ~\ln{\left[~ \frac{G(1)}{G(\infty)} \right]~}.
\label{a_sing_D_gt_0}
\end{equation}
It is straightforward to verify that $t_{*}$ once again corresponds to a Big Rip singularity.

\textbf{Case $\Lambda = 0$.} We illustrate the above asymptotic discussion with exact analytical solutions for the case of a null CC. The diffusion function is described by  $Q_{0}<0$ and $\beta <0$. As was mentioned above, we also have the constraint $\beta < 3 \gamma$. 

In this case, the Hubble parameter takes the form
\begin{equation}
H(a)= \left[A ~ a^{3|\gamma|} + B~ a^{|\beta|} \right]^{1/2},
\label{Hubble_a1}
\end{equation}
with $A$ and $B$ defined, respectively, by 
\begin{equation}
A = \frac{\rho_{0}}{3} ~\left(1 - \frac{|\beta| |Q_{0}| / \rho_{0}}{|\beta|-3|\gamma| } \right) \quad \text{and} \quad B = \frac{|\gamma| ~| Q_{0}|}{|\beta|-3|\gamma|} .
\label{def_II_A_B}
\end{equation}

In terms of the scale factor we can integrate the above expression which yields the following integrals
\begin{equation} \label{integraltype}
\int_{a_0}^{a} \frac{da}{\sqrt{Aa^{3
|\gamma|+2} + Ba^{|\beta|+2}}} = \int_{t_0}^{t} dt.
\end{equation} 
There are particular situations in which choosing specific values for $|\gamma|$ and $|\beta|$ makes it possible to express the integral in terms of rational functions. 

\textbf{Big Rip singularity for the particular values $3\gamma = -1$, and  $\beta = -2$.} For this case, we have  
\begin{equation} \label{case12}
\int \frac{da}{a\sqrt{Aa + Ba^{2}}} = -\frac{2\sqrt{Aa + Ba^{2}}}{Aa},
\end{equation} 

Taking the initial condition $a_0=1$ we find that the scale factor is given by
which gives the following expression for the scale factor as a function of the cosmic time
\begin{equation} \label{solutionfoat}
a(t) = \frac{A}{\left [\sqrt{A+B}-\frac{A(t-t_0)}{2} \right ]^2-B},
\end{equation}  
which indicates that for this election the universe evolves to a Big Rip singularity in a finite time $t=t_{BR}$ given by 
\begin{equation} \label{solutionBR}
t_{BR}-t_0 = 2\frac{\left (\sqrt{A+B}\mp\sqrt{B} \right )}{A}.
\end{equation}
However, since the scale parameter a must be non-negative, a straightforward analysis reveals that the appropriate solution is the earlier one, corresponding to the negative sign in the expression above.

In addition, for the particular case $B \geq A $, the above equation up to first order goes into
\begin{equation} \label{solution_approx_BR}
t_{BR}-t_0 \approx \frac{1}{\sqrt{B}},
\end{equation}
which exactly coincides with the first order of the general perturbative result obtained in Eq. (\ref{BR_first_order}), when the particular values $\beta = -2$ and $3\gamma = -1$ are used.

The above result is fully consistent with the general perturbative analysis, which demonstrates that throughout the entire parameter region $\beta < 3 \gamma < 0$, Big Rip singularities inevitably arise, regardless of the specific value of the exponential coefficient $\beta$ used to model the power-law behavior of the energy diffusion function.

Moreover, we can evaluate the coefficients $A$ and $B$ in terms of the model parameters $\rho_{0}$ and $Q_{0}$, obtaining  
\begin{equation} \label{solutionBR1}
t_{BR}-t_0 = 2\sqrt{3}\frac{\left (\sqrt{\rho_{0}-|Q_{0}|}-\sqrt{|Q_{0}|} \right )}{\rho_{0}-|Q_{0}|}.
\end{equation}
Furthermore, defining the dimensionless densities $\Omega_{0m}=\rho_{0m}/3H_{0}$ and $\Omega_{0Q}=|Q_{0}|/3H_{0}$, Eq. (\ref{solutionBR1}) takes the following form for $t_{BR}$
\begin{equation} \label{solutionforn}
t_{BR}= t_0 + \frac{1}{H_{0}}\frac{\left (\sqrt{\Omega_{0m}-\Omega_{0Q}}-\sqrt{\Omega_{0Q}} \right )}{\Omega_{0m}-\Omega_{0Q}}.
\end{equation}     
If $\Omega_{0m} \gg \Omega_{0Q}$, we can expand to the first order, obtaining
\begin{equation}
t \approx \frac{1}{H_0\sqrt{\Omega_{0m}}} \left(1 + \frac{\Omega_{0Q}}{\Omega_{0m}} - \frac{3}{2}\sqrt{\frac{\Omega_{0Q}}{\Omega_{0m}}}\right).
\end{equation}
The Hubble parameter and its derivative $\dot{H}$ corresponding to the solution given in Eq. (\ref{solutionfoat}) are
\begin{equation}
H=\frac{\dot{a}(t)}{a(t)} = \frac{A\left(\sqrt{A+B} - \frac{A}{2}(t-t_0)\right)}{\left(\sqrt{A+B} - \frac{A}{2}(t-t_0)\right)^2 - B},
\end{equation}
and 
\begin{equation}
\dot{H}= \frac{A^2\left[\left(\sqrt{A+B} - \frac{A}{2}(t-t_0)\right)^2 + B\right]}{2\left[\left(\sqrt{A+B} - \frac{A}{2}(t-t_0)\right)^2 - B\right]^2}
\end{equation}
respectively, showing that both parameters diverge at $t=t_{BR}$ as well. Notice that, since $p=(\gamma -1)\rho$, the density and pressure go to infinity at $t_{BR}$ as well. 

\textbf{Big Rip singularity for the particular values $3\gamma = -2$, and  $\beta = -3$.} An additional analytically solvable configuration leading to Big Rip singularities is obtained when the model parameters are $\Lambda = 0$, $3\gamma = -2$, and  $\beta = -3$. For this case, the integral of Eq.~(\ref{scale_parameter}), up to an integration constant, yields 
\begin{equation} \label{integral2and3}
\int \frac{da/ a}{H(a)} = -\frac{\sqrt{A + Ba}}{A~ a} + \frac{B}{A^{3/2}} \tanh^{-1} \left(\frac{\sqrt{A}}{\sqrt{A+Ba}}\right)
\end{equation}

In the present scenario, the scale parameter $a$ is not bound from below and therefore can use $a_{0} =1$ as the lower integration limit to evaluate the above integral. It yields

\begin{eqnarray}
t-t_{0} &=& \frac{\sqrt{A + B}}{A} - \frac{\sqrt{A + Ba}}{A~ a} + \frac{B}{A^{3/2}} ~\times \nonumber\\
        & &  \left[ \tanh^{-1} \left(\frac{\sqrt{A}}{\sqrt{A+Ba}}\right) -  \tanh^{-1} \left(\frac{\sqrt{A}}{\sqrt{A+B}}\right) \right].
\nonumber
\end{eqnarray}
 Since $\tanh^{-1}(x)$ is a surjective function in $[-1, 1]$, even if the above expression cannot be analytically inverted to obtain $a(t)$, since the argument $x = \sqrt{A / (A+Ba)}$ lies in the domain for all $a \in \mathbb{R}_{+} \cup \{ 0 \}$, there always exists the limit when $a \rightarrow \infty$, leading to the future finite singular time $t_{s} $ :
\begin{equation} 
t_{s} - t_{0} = \frac{\sqrt{A + B}}{A } - \frac{B}{A^{3/2}} ~\tanh^{-1} \left(\sqrt{ \frac{A}{A+B}}\right) 
\label{singular_t}
\end{equation} 
Therefore, at $t = t_{s}$ the scale parameter diverges. Moreover, since both the Hubble parameter and its time derivative are given by $H(a) = \sqrt{Aa^{2} + Ba^{3}}$ and $\dot{H} = Aa^{2}+3Ba^{3}/2$, respectively, both quantities also diverge at this time. Finally, from Eq.~(\ref{rho_m_a}), it follows that the matter energy density $\rho$ and the absolute value of the pressure $|p| = |\gamma -1|~\rho$ also blow up. In other words, $t_{s}$, corresponds to a Big Rip singularity, according to the classification introduced in Section $IV$.

\textbf{Case $\Lambda < 0$}. Finally, for $\Lambda < 0$, the discriminant of the quadratic polynomial $Ba^2 + Aa + \Lambda/3$ appearing in the denominator of Eq.~(\ref{part_b_g_Lambda}) is positive, Nevertheless, the requirement that $H^2(a)$ remains non-negative imposes a constraint on the scale factor $a$, which, when expressed in terms of the variable $u = 1/a $ can be written as

\begin{equation}
u \in [0, u_{+}], ~ \text{with} ~ u_{\pm} = \frac{3A}{2|\Lambda|} \pm \sqrt{ \left( \frac{3A}{2 \Lambda} \right)^2 + \frac{3B}{|\Lambda|}}
\label{limits_on_a_D_gt_0a}
\end{equation} 
The upper bound for $u$ corresponds to a minimum value of the scale parameter, $a \leq a_{m}$, with $a_{m} = 1/u_{+}$. 

Keeping this in mind, the result for the integral of Eq.~(\ref{part_b_g_Lambda}) can be written in the suitable representation

\begin{equation}
(\tau-\tau_{m}) = \sin^{-1} \left( \frac{1/a_{m} - u_{\triangle}}{D} \right) -  \sin^{-1} \left( \frac{1/a - u_{\triangle}}{D} \right),
\label{t_BR_for_part_b_g_Lambda_lt_0}
\end{equation}
where we have introduced the following definitions: $u_{\triangle} = 3A/2 |\Lambda|$, $a_{m} = a(t_{m})$, $t = \tau~\sqrt{3/|\Lambda|}$, and $D $ is the square root of the discriminant, i.e., $D = \sqrt{(3A/2 \Lambda)^2 + 3B / |\Lambda|}$.

The above expression can be solved for the scale parameter yielding

\begin{equation}
a(t) = \frac{1}{ u_{\triangle} + D~\sin \left[\sin^{-1}\left(\frac{1/a_{m} - u_{\triangle} }{D} \right) - (\tau-\tau_{m}) \right] }.
\label{a_for_part_b_g_Lambda_lt_0}
\end{equation}  
As explained above, the existence of an upper bound for the inverse variable $u = 1/a$ determined from Eq.~(\ref{limits_on_a_D_gt_0a}) leads to the existence of a lower bound for the scale factor $a$, which can be written as

\begin{equation}
a(t) \geq \frac{|\Lambda|}{3A} \left( 1 - \frac{B |\Lambda|}{3A^2} + ... ~\right)
\approx \frac{|\Lambda|}{3A}
\label{lower_bound_a_Lambda_lt_o} 
\end{equation}  
where the last approximation holds for relatively small values of the cosmological constant, which is a reasonable assumption considering the current cosmological observations for $\Lambda$. 

Now, the scale parameter can be cast in the simple analytical form 

\begin{equation}
a(t) = \frac{1}{ u_{\triangle} + D~\sin \left[\frac{\pi}{2} - \sqrt{\frac{|\Lambda|}{3}}~(t-t_{m}) \right] }.
\label{a_sing_for_part_b_g_Lambda_lt_0}
\end{equation}  
Since $u_{\triangle} < D$, there always exists a value for $t-t_{m}$ at which the denominator of the above expression vanishes. This value corresponds to the singularity of $a$ at
\begin{equation}
t_{s} - t_{m} = \sqrt{\frac{3}{|\Lambda|}} ~\biggl[~\pi/2 + \sin^{-1}\left(\frac{u_{\triangle}}{D}\right) \biggr].  
\label{singularity_Lambda_lt_0}
\end{equation}
Again, the above equation for relatively small values of the cosmological constant goes into

\begin{equation}
t_{s} -t_{m} \approx \sqrt{\frac{3}{|\Lambda|}} \left( \pi - \frac{2}{A}~\sqrt{\frac{B~|\Lambda|}{3}} + ~... \right)
\label{dt_singular_Lambda_lt_0}
\end{equation}
Since the scale factor $a$ diverges for $t=t_{s}$, and consequently both $H$ and $\rho$ also diverge (as can be easily verified), $t_{s}$ corresponds to a Big Rip–type singularity.

A final remark is in order. When comparing the above expression with Eq.~(\ref{BR_time}), which gives the Big Rip time for the particular case where the coefficient $A$ vanishes, one finds a factor of two difference when the value $\beta = -2$ is inserted into the latter expression. This apparent discrepancy is readily understood: the computation leading to Eq.~(\ref{BR_time}) evaluates the time required to reach the singularity starting from the minimum scale factor $a_{m}^{p} = (|\Lambda| / 3B)^{1/|\beta|}$, whereas the minimum value of $a$ in the more general case described by Eq.~(\ref{dt_singular_Lambda_lt_0}) differs from this point as

\begin{equation}
a_{m} = \left( \frac{3A}{2|\Lambda|} + \sqrt{ \left( \frac{3A}{2|\Lambda|} \right)^2 + \frac{3B}{|\Lambda|}} ~\right)^{-1}.
\end{equation}
This shows that the two results are equivalent for $A = 0$, however, in the general case they are not intrinsically comparable.
 
\section{Final Remarks}
\label{sec:final}

In this work, we have investigated the appearance of future singularities within the UG framework, analyzing the future evolution of universes containing a single fluid characterized by a barotropic EoS, $p=(\gamma -1)\rho$, where $\gamma$ can account for both phantom and non phantom regimes. Our goal was to gain a more profound insight into the physical scenarios that this theory is capable of producing.

We introduce an energy diffusion function (EDF), $Q(z)$, through a phenomenological Ansatz, and impose the condition of positive entropy production associated with the intrinsically non-adiabatic evolution of the fluid. This requirement translates into the constraint $\frac{dQ}{dz} > 0$, or equivalently, $\frac{dQ}{dt} < 0$. Consequently, within UG, the second law of thermodynamics restricts the EDF to functioning solely as an energy source for the cosmic fluid and never as an energy sink.

To ensure that the thermodynamic condition outlined above is satisfied, we adopt an Ansatz in which $Q$ is taken to be a monotonically decreasing function of cosmic time—or, equivalently, a monotonically increasing function of redshift. This behavior is straightforwardly implemented by postulating a power-law dependence on redshift, $Q \sim (1+z)^{\beta}$, where the exponent $\beta$ determines the temporal evolution of the diffusion function.

In the framework of General Relativity, the positivity of the fluid energy density is ensured from the outset, since the chosen phantom equation of state already satisfies this requirement. In contrast, within UG, we have shown that the contribution of $Q$ to the energy density $\rho>0$ imposes stringent constraints on the model parameters in order to guaranty throughout the entire cosmic evolution (see, for instance, Eq.~(\ref{eq:energy3})).

Using the parameter constraints discussed in the previous sections, we have found that if the main fluid has a non-phantom EoS, no future singularity arises and the Universe approaches the usual de Sitter phase at late times, provided that $\Lambda >0$. In this sense, we have derived a complementary no-go theorem: within the thermodynamically admissible sector of unimodular gravity—characterized by positive entropy production—a diffusion term alone cannot generate Big Rip singularities when the matter component is non-phantom and $\Lambda > 0$. 

To the best of our knowledge, this constitutes the first general exclusion of diffusion-driven Big Rip singularities under physically motivated thermodynamic constraints.

In contrast, for $\Lambda<0$ a Big Crunch singularity may occur, and—more interestingly—a Big Rip singularity can also emerge for a specific choice of the initial energy density of the fluid and of the EDF, $Q$. This novel result shows that the diffusion function can induce an effective phantom behavior in the matter sector, leading to a future singularity of this type, even in the presence of a negative cosmological constant, despite the fact that the main fluid itself satisfies a non-phantom "bare" EoS. This reveals a novel mechanism for the emergence of future singularities, which bears some analogy to Big Rip solutions found for dissipative fluids within Eckart’s theory in standard General Relativity with $\Lambda < 0$ \cite{cruz2022singularities}.

Although our findings depend on the specific choices made in this preliminary analysis, we have nonetheless demonstrated that enforcing positive entropy production, along with conditions $H^2(z) > 0$ and a positive energy density, places strong restrictions on the range of viable late-time evolutions. In fact, within the physically allowed parameter region, we find that if the fluid is non-phantom, Big Rip singularities are excluded—except in the presence of a negative cosmological constant, and only for a very specific choice of the initial energy density $\rho_{0}$ (see Eq.~(\ref{thermo_law})). These results, which closely mirror the situation in General Relativity, arise from the sign restrictions on $Q_{0}$ and $\beta$ imposed by thermodynamic consistency. In this regard, the inherent non-adiabatic character of the UG framework requires that a consistent thermodynamic evolution be established first, before investigating any potential future-singularity scenarios. 

\section*{Acknowledgments}
NC, SL and GP acknowledge financial support from the Chilean National Agency for Research and Development (ANID) through Fondecyt Grant No. 1250969,
Chile. MC work has been partially supported by S.N.I.I. (SECIHTI-M\'exico).

\bibliographystyle{ieeetr}
\bibliography{biblio.bib}

@article{Einstein:1919gv,
    author = "Einstein, Albert",
    title = "{Spielen Gravitationsfelder im Aufbau der materiellen Elementarteilchen eine wesentliche Rolle?}",
    journal = "Sitzungsber. Preuss. Akad. Wiss. Berlin (Math. Phys. )",
    volume = "1919",
    pages = "349--356",
    year = "1919"
}

@article{BUCHMULLER1988292,
title = "{Einstein gravity from restricted coordinate invariance}",
journal = {Physics Letters B},
volume = {207},
number = {3},
pages = {292-294},
year = {1988},
issn = {0370-2693},
doi = {https://doi.org/10.1016/0370-2693(88)90577-1},
url = {https://www.sciencedirect.com/science/article/pii/0370269388905771},
author = {W. Buchmüller and N. Dragon}
}

@article{PhysRevD.40.1048,
  title = "{Unimodular theory of canonical quantum gravity}",
  author = {Unruh, W. G.},
  journal = {Phys. Rev. D},
  volume = {40},
  issue = {4},
  pages = {1048--1052},
  numpages = {0},
  year = {1989},
  month = {Aug},
  publisher = {American Physical Society},
  doi = {10.1103/PhysRevD.40.1048},
  url = {https://link.aps.org/doi/10.1103/PhysRevD.40.1048}
}

@article{10.1063/1.529283,
    author = {Ng, Y. Jack and van Dam, H.},
    title = "{Unimodular theory of gravity and the cosmological constant}",
    journal = {Journal of Mathematical Physics},
    volume = {32},
    number = {5},
    pages = {1337-1340},
    year = {1991},
    month = {05},
    issn = {0022-2488},
    doi = {10.1063/1.529283},
    url = {https://doi.org/10.1063/1.529283},
    eprint = {https://pubs.aip.org/aip/jmp/article-pdf/32/5/1337/19170868/1337\_1\_online.pdf},
}

@article{10.1063/1.1328077,
    author = {Finkelstein, David R. and Galiautdinov, Andrei A. and Baugh, James E.},
    title = "{Unimodular relativity and cosmological constant}",
    journal = {Journal of Mathematical Physics},
    volume = {42},
    number = {1},
    pages = {340-346},
    year = {2001},
    month = {01},
    issn = {0022-2488},
    doi = {10.1063/1.1328077},
    url = {https://doi.org/10.1063/1.1328077},
    eprint = {https://pubs.aip.org/aip/jmp/article-pdf/42/1/340/19287338/340\_1\_online.pdf},
}

@article{Ellis_2011,
doi = {10.1088/0264-9381/28/22/225007},
url = {https://dx.doi.org/10.1088/0264-9381/28/22/225007},
year = {2011},
month = {oct},
publisher = {IOP Publishing},
volume = {28},
number = {22},
pages = {225007},
author = {Ellis, George F R and van Elst, Henk and Murugan, Jeff and Uzan, Jean-Philippe},
title = "{On the trace-free Einstein equations as a viable alternative to general relativity}",
journal = {Classical and Quantum Gravity}
}

@article{ellis_trace-free_2013,
	title = {The trace-free {Einstein} equations and inflation},
	volume = {46},
	issn = {1572-9532},
	url = {https://doi.org/10.1007/s10714-013-1619-5},
	doi = {10.1007/s10714-013-1619-5},
	number = {1},
	journal = {General Relativity and Gravitation},
	author = {Ellis, George F. R.},
	month = dec,
	year = {2013},
	pages = {1619},
}

@article{Pearle:1976ka,
      author         = "Pearle, Philip M.",
      title          = "{Reduction of the State Vector by a Nonlinear Schrodinger
                        Equation}",
      journal        = "Phys. Rev.",
      volume         = "D13",
      year           = "1976",
      pages          = "857-868",
      doi            = "10.1103/PhysRevD.13.857",
      SLACcitation   = "%%CITATION = PHRVA,D13,857;%%"
}

@article{Ghirardi:1985mt,
      author         = "Ghirardi, G. C. and Rimini, A. and Weber, T.",
      title          = "{A Unified Dynamics for Micro and MACRO Systems}",
      journal        = "Phys. Rev.",
      volume         = "D34",
      year           = "1986",
      pages          = "470",
      doi            = "10.1103/PhysRevD.34.470",
      reportNumber   = "IC/85/292",
      SLACcitation   = "%%CITATION = PHRVA,D34,470;%%"
}

@article{Pearle:1988uh,
      author         = "Pearle, Philip M.",
      title          = "{Combining Stochastic Dynamical State Vector Reduction
                        With Spontaneous Localization}",
      journal        = "Phys. Rev.",
      volume         = "A39",
      year           = "1989",
      pages          = "2277-2289",
      doi            = "10.1103/PhysRevA.39.2277",
      reportNumber   = "IC/88/99",
      SLACcitation   = "%%CITATION = PHRVA,A39,2277;%%"
}

@article{Ghirardi:1989cn,
      author         = "Ghirardi, Gian Carlo and Pearle, Philip M. and Rimini,
                        Alberto",
      title          = "{Markov Processes in Hilbert Space and Continuous
                        Spontaneous Localization of Systems of Identical
                        Particles}",
      journal        = "Phys. Rev.",
      volume         = "A42",
      year           = "1990",
      pages          = "78-79",
      doi            = "10.1103/PhysRevA.42.78",
      reportNumber   = "IC-89-44",
      SLACcitation   = "%%CITATION = PHRVA,A42,78;%%"
}

@article{perez2019dark,
  title={Dark energy from quantum gravity discreteness},
  author={Perez, Alejandro and Sudarsky, Daniel},
  journal={Physical review letters},
  volume={122},
  number={22},
  pages={221302},
  year={2019},
  publisher={APS}
}

@article{efstathiou2025challenges,
  title="{Challenges to the $\Lambda$CDM cosmology}",
  author={Efstathiou, George},
  journal={Philosophical Transactions A},
  volume={383},
  number={2290},
  pages={20240022},
  year={2025},
  publisher={The Royal Society}
}

@article{perez2021resolving,
  title={Resolving the H 0 tension with diffusion},
  author={Perez, Alejandro and Sudarsky, Daniel and Wilson-Ewing, Edward},
  journal={General Relativity and Gravitation},
  volume={53},
  number={1},
  pages={7},
  year={2021},
  publisher={Springer}
}

@article{LinaresCedeno:2020uxx,
    author = "Linares Cede\~no, Francisco X. and Nucamendi, Ulises",
    title = "{Revisiting cosmological diffusion models in Unimodular Gravity and the $H_0$ tension}",
    doi = "10.1016/j.dark.2021.100807",
    journal = "Phys. Dark Univ.",
    volume = "32",
    pages = "100807",
    year = "2021"
}

@article{Corral:2020lxt,
    author = "Corral, Crist\'bal and Cruz, Norman and Gonz\'alez, Esteban",
    title = "{Diffusion in unimodular gravity: Analytical solutions, late-time acceleration, and cosmological constraints}",
    journal = "Phys. Rev. D",
    volume = "102",
    pages = "023508",
    year = "2020",
    doi = "10.1103/PhysRevD.102.023508", 
    archivePrefix = "arXiv",
    primaryClass = "gr-qc"
    
}

@article{Riess_2018,
	doi = {10.3847/1538-4357/aaadb7},
	url = {https://doi.org/10.3847%2F1538-4357%2Faaadb7},
	year = 2018,
	month = {mar},
	publisher = {American Astronomical Society},
	volume = {855},
	number = {2},
	pages = {136},
	author = {Adam G. Riess and others},
	title = {New Parallaxes of Galactic Cepheids from Spatially Scanning {theHubble} Space Telescope: Implications for the Hubble Constant},
	journal = {The Astrophysical Journal},
}

@ARTICLE{2019NatAs...3..272R,
       author = {{Risaliti}, G. and {Lusso}, E.},
        title = "{Cosmological Constraints from the Hubble Diagram of Quasars at High Redshifts}",
      journal = {Nature Astronomy},
         year = 2019,
        month = jan,
       volume = {3},
        pages = {272-277},
          doi = {10.1038/s41550-018-0657-z},
       adsurl = {https://ui.adsabs.harvard.edu/abs/2019NatAs...3..272R}
}

@article{cruz_exploring_2024,
	title = {Exploring thermodynamics inconsistencies in unimodular gravity: a comparative study of two energy diffusion functions},
	volume = {84},
	issn = {1434-6052},
	shorttitle = {Exploring thermodynamics inconsistencies in unimodular gravity},
	url = {https://link.springer.com/10.1140/epjc/s10052-024-13523-w},
	doi = {10.1140/epjc/s10052-024-13523-w},
	language = {en},
	number = {11},
	urldate = {2025-06-07},
	journal = {The European Physical Journal C},
	author = {Cruz, Miguel and Cruz, Norman and Lepe, Samuel},
	month = nov,
	year = {2024},
	pages = {1186},
}

@article{PhysRevD.99.123525,
  title = "{Cosmic acceleration in unimodular gravity}",
  author = "{Garc\'{\i}a-Aspeitia, Miguel A., et al}",
  journal = {Phys. Rev. D},
  volume = {99},
  issue = {12},
  pages = {123525},
  numpages = {7},
  year = {2019},
  month = {Jun},
  publisher = {American Physical Society},
  doi = {10.1103/PhysRevD.99.123525},
  url = {https://link.aps.org/doi/10.1103/PhysRevD.99.123525}
}

@book{Callen:450289,
      author        = "Callen, Herbert B",
      title         = "{Thermodynamics and an introduction to thermostatistics;
                       2nd ed.}",
      publisher     = "Wiley",
      address       = "New York, NY",
      year          = "1985",
      url           = "https://cds.cern.ch/record/450289",
}

@article{bousis2024hubble,
  title="{Hubble tension tomography: BAO vs SN Ia distance tension}",
  author={Bousis, Dimitrios and Perivolaropoulos, Leandros},
  journal={Physical Review D},
  volume={110},
  number={10},
  pages={103546},
  year={2024},
  publisher={APS}
}

@article{yang2024quintom,
  title="{Quintom cosmology and modified gravity after DESI 2024}",
  author={Yang, Yuhang and Ren, Xin and Wang, Qingqing and Lu, Zhiyu and Zhang, Dongdong and Cai, Yi-Fu and Saridakis, Emmanuel N},
  journal={Science Bulletin},
  volume={69},
  number={17},
  pages={2698--2704},
  year={2024},
  publisher={Elsevier}
}

@article{giare2024interacting,
  title="{Interacting dark energy after DESI baryon acoustic oscillation measurements}",
  author={Giar{\`e}, William and Sabogal, Miguel A and Nunes, Rafael C and Di Valentino, Eleonora},
  journal={Physical Review Letters},
  volume={133},
  number={25},
  pages={251003},
  year={2024},
  publisher={APS}
}

@article{colgain2024does,
  title="{Does DESI 2024 Confirm $\Lambda$CDM?}",
  author={Colg{\'a}in, Eoin {\'O} and Dainotti, Maria Giovanna and Capozziello, Salvatore and Pourojaghi, Saeed and Sheikh-Jabbari, MM and Stojkovic, Dejan},
  journal={arXiv preprint arXiv:2404.08633},
  year={2024}
}

@article{park2024using,
  title="{Using non-DESI data to confirm and strengthen the DESI 2024 spatially flat w 0 wa CDM cosmological parametrization result}",
  author={Park, Chan-Gyung and P{\'e}rez, Javier de Cruz and Ratra, Bharat},
  journal={Physical Review D},
  volume={110},
  number={12},
  pages={123533},
  year={2024},
  publisher={APS}
}

@article{FRAMPTON2012204,
title = {Models for little rip dark energy},
journal = {Physics Letters B},
volume = {708},
number = {1},
pages = {204-211},
year = {2012},
issn = {0370-2693},
doi = {https://doi.org/10.1016/j.physletb.2012.01.048},
url = {https://www.sciencedirect.com/science/article/pii/S0370269312000767},
author = {Paul H. Frampton and Kevin J. Ludwick and Shinʼichi Nojiri and Sergei D. Odintsov and Robert J. Scherrer}
}

@article{Blanchard_2024,
   title="{ΛCDM is alive and well}",
   volume={7},
   ISSN={2565-6120},
   url={http://dx.doi.org/10.33232/001c.117170},
   DOI={10.33232/001c.117170},
   journal={The Open Journal of Astrophysics},
   publisher={Maynooth University},
   author={Blanchard, Alain and Héloret, Jean-Yves and Ilić, Stéphane and Lamine, Brahim and Tutusaus, Isaac},
   year={2024},
   month=may 
}

@article{Visinelli_neg_CC_2019,
   title="{Revisiting a Negative Cosmological Constant from Low-Redshift Data}",
   volume={11:1035},
   ISSN={2073-8994},
   url={https://www.mdpi.com/2073-8994/11/8/1035},
   DOI={10.3390/sym11081035},
   journal={Symmetry},
   publisher={MDPI Open Access Journals},
      author={Visinelli, L. and Vagnozzi, S. and Danielsson, U.},
   year={2019},
   month=aug 
}

@Article{tiposdeBigRip,
  author  = {Shin’ichi Nojiri and Sergei D. Odintsov and Shinji Tsujikawa},
  journal = {Physical Review D},
  title   = "{Properties of singularities in (phantom) dark energy universe}",  
  year    = {2005},
  volume={71},
  pages={063004}
}

@Article{clasificacionBigRipdetallada,
  author = {Mariam Bouhmadi-López, Claus Kiefer, Prado Martín-Moruno},
  journal = {General Relativity and Gravitation 51 (2019) 10, 135},
  title = "{Phantom singularities and their quantum fate: general relativity and beyond—a CANTATA COST action topic}",
  year = {2019},
  volume = {51},
  pages = {135}
}

@article{caldwell2003phantom,
  title="{Phantom energy: dark energy with $w<-1$ causes a cosmic doomsday}",
  author={Caldwell, Robert R and Kamionkowski, Marc and Weinberg, Nevin N},
  journal={Physical Review letters},
  volume={91},
  number={7},
  pages={071301},
  year={2003},
  publisher={APS}
}

@article{gonzalez2004k,
  title="{K-essential phantom energy: Doomsday around the corner?}",
  author={Gonz{\'a}lez-D{\'i}az, Pedro F},
  journal={Physics Letters B},
  volume={586},
  number={1-2},
  pages={1--4},
  year={2004},
  publisher={Elsevier}
}

@article{cruz2022singularities,
  title="{Singularities and soft-Big Bang in a viscous $\Lambda$ CDM model}",
  author={Cruz, Norman and Gonz{\'a}lez, Esteban and Jovel, Jose},
  journal={Physical Review D},
  volume={105},
  number={2},
  pages={024047},
  year={2022},
  publisher={APS}
}

@article{frampton2011little,
  title={The little rip},
  author={Frampton, Paul H and Ludwick, Kevin J and Scherrer, Robert J},
  journal={Physical Review D—Particles, Fields, Gravitation, and Cosmology},
  volume={84},
  number={6},
  pages={063003},
  year={2011},
  publisher={APS}
}

@article{brevik2011viscous,
  title={Viscous little rip cosmology},
  author={Brevik, I and Elizalde, E and Nojiri, S’i and Odintsov, SD},
  journal={Physical Review D—Particles, Fields, Gravitation, and Cosmology},
  volume={84},
  number={10},
  pages={103508},
  year={2011},
  publisher={APS}
}

@article{bouhmadi2013tradeoff,
  title="{Tradeoff between smoother and sooner “little rip”}",
  author={Bouhmadi-L{\'o}pez, Mariam and Chen, Pisin and Liu, Yen-Wei},
  journal={The European Physical Journal C},
  volume={73},
  number={9},
  pages={2546},
  year={2013},
  publisher={Springer}
}

@article{albarran2016classical,
  title="{Classical and quantum cosmology of the little rip abrupt event}",
  author={Albarran, Imanol and Bouhmadi-L{\'o}pez, Mariam and Kiefer, Claus and Marto, Jo{\~a}o and Vargas Moniz, Paulo},
  journal={Physical Review D},
  volume={94},
  number={6},
  pages={063536},
  year={2016},
  publisher={APS}
}

@article{calderon2021negative,
  title="{Negative cosmological constant in the dark sector?}",
  author={Calder{\'o}n, Rodrigo and Gannouji, Radouane and L’Huillier, Benjamin and Polarski, David},
  journal={Physical Review D},
  volume={103},
  number={2},
  pages={023526},
  year={2021},
  publisher={APS}
}

@article{akarsu2020graduated,
  title="{Graduated dark energy: Observational hints of a spontaneous sign switch in the cosmological constant}",
  author={Akarsu, {\"O}zg{\"u}r and Barrow, John D and Escamilla, Luis A and Vazquez, J Alberto},
  journal={Physical Review D},
  volume={101},
  number={6},
  pages={063528},
  year={2020},
  publisher={APS}
}
\end{document}